\documentclass[fleqn,usenatbib]{mnras}

\usepackage{newtxtext,newtxmath}
\usepackage[T1]{fontenc}
\usepackage{ae,aecompl}
\usepackage{longtable}

\usepackage{graphicx}	% Including figure files
\usepackage{amsmath}	% Advanced maths commands
\usepackage{amssymb}	% Extra maths symbols

\newcommand{\ms}{\mbox{m\,s$^{-1}$}}
\newcommand{\kms}{\mbox{km\,s$^{-1}$}}
\newcommand{\Msun}{\mbox{$M_{\odot}$}}

\newcommand{\Mjup}{\mbox{$M_{\rm Jup}$}}

\newcommand{\gtsimeq}{\raisebox{-0.6ex}{$\,\stackrel
        {\raisebox{-.2ex}{$\textstyle >$}}{\sim}\,$}}

\title[PPPS VIII]{The Pan-Pacific Planet Search. VIII. Complete results and the occurrence rate of planets around low-luminosity giants. }

\author[R.A. Wittenmyer et al.]{
Robert A. Wittenmyer,$^{1}$\thanks{E-mail: rob.w@usq.edu.au}
R.P. Butler,$^{2}$
Jonathan Horner,$^{1}$
Jake Clark,$^{1}$
C.G. Tinney,$^{3}$
\newauthor B.D. Carter,$^{1}$
Liang Wang,$^{4}$
John Asher Johnson,$^{5}$
Michaela Collins$^{6}$
\\
$^{1}$University of Southern Queensland, Centre for Astrophysics, USQ Toowoomba, QLD 4350 Australia \\
$^{2}$Department of Terrestrial Magnetism, Carnegie Institution of Washington, 5241 Broad Branch Road, NW, Washington, DC 20015-1305, USA \\
$^{3}$School of Physics and Australian Centre for Astrobiology, University of New South Wales, Sydney 2052, Australia \\
$^{4}$Nanjing Institute of Astronomical Optics and Technology, Chinese Academy of Sciences, Bancang Street 188, Nanjing, China 210042 \\
$^{5}$Harvard-Smithsonian Center for Astrophysics, 60 Garden Street, Cambridge, MA 02138, USA \\
$^{6}$Department of Physical Sciences, Kutztown University, Kutztown, PA 19530, USA
}

\date{Accepted XXX. Received YYY; in original form ZZZ}

\pubyear{2019}

\begin{document}
\label{firstpage}
\pagerange{\pageref{firstpage}--\pageref{lastpage}}
\maketitle

% Abstract of the paper
\begin{abstract}
Our knowledge of the populations and occurrence rates of planets orbiting evolved intermediate-mass stars lags behind that for solar-type stars by at least a decade.  Some radial velocity surveys have targeted these low-luminosity giant stars, providing some insights into the properties of their planetary systems.  Here we present the final data release of the Pan-Pacific Planet Search, a 5-year radial velocity survey using the 3.9m Anglo-Australian Telescope.  We present 1293 precise radial velocity measurements for 129 stars, and highlight six potential substellar-mass companions which require additional observations to confirm.  Correcting for the substantial incompleteness in the sample, we estimate the occurrence rate of giant planets orbiting low-luminosity giant stars to be approximately 7.8$^{+9.1}_{-3.3}$\%.  This result is consistent with the frequency of such planets found to orbit main-sequence A-type stars, from which the PPPS stars have evolved. 

%In which our heroes dump all the RVs and hand wave some occurrence rates with limited data.
\end{abstract}

% Select between one and six entries from the list of approved keywords.
% Don't make up new ones.
\begin{keywords}
techniques: radial velocities -- planets and satellites: detection
\end{keywords}

%%%%%%%%%%%%%%%%%%%%%%%%%%%%%%%%%%%%%%%%%%%%%%%%%%

%%%%%%%%%%%%%%%%% BODY OF PAPER %%%%%%%%%%%%%%%%%%

\section{Introduction}

With the discovery of the first planets orbiting other stars \citep[e.g.][]{campbell88,lathamsworld,51peg}, astronomers gained our first insight into the degree to which the Solar system is unique.  In the three decades since, the global search for exoplanets has led to the discovery of more than 4000 planets orbiting nearby stars.  Those discoveries have revealed the diversity and ubiquity of planetary systems - with the great majority of systems discovered proving to be remarkably different to the Solar system \citep[e.g.][]{petigura13, winn15, bryan19}. 

The last decade has seen the dawn of the golden age of space-based transit discoveries, which has led to the number of known exoplanets climbing by more than an order of magnitude. As a result, more than 80\% of all currently confirmed exoplanets were first identified by the \textit{Kepler} and \textit{TESS} space telescopes \citep{borucki10, ricker15}.  Those missions have been wildly successful in expanding our understanding of planetary system properties and architectures \citep[e.g.][]{lissauer11, lissauer14, raymond18, zhu19} and, together with long-running radial velocity survey programs, have allowed us to study the occurrence rate of planets around Solar-type and late-type stars \citep[e.g.][]{endl06, otoole09, fressin13, h19}.  This has, in turn, opened a window on the planet formation history of the galaxy, and has allowed conclusions to be drawn on the occurrence rate of true Solar system analogues ($\sim 24\%$ of planetary systems contain Earth-like planets; e.g. \citealt{barbato18}, and $\sim$3-6\% contain Jupiter analogues; e.g. \citealt{zech13}, \citealt{newjupiters}, \citealt{agnew18}, and \citealt{borg19}).

Whilst the situation for Solar-type stars is now relatively understood, our knowledge of the occurrence and nature of planets around evolved stars remains relatively stunted.  The main reason for this is that transit surveys intentionally bias against targeting stars that may be evolved.  Such stars have larger radii, and hence the signal that results from planetary transits will be correspondingly diluted.  The \textit{Kepler} and \textit{TESS} prime target lists selected against giant stars \citep[e.g.][]{KIC, stassun18}, though the \textit{TESS} full frame images are a valuable bias-free source of transit photometry for all types of stars.  The confirmation of planet candidates transiting evolved stars, however, is frustrated not only by the smaller size of the signal, but also by intrinsic stellar variability, resulting in a high rate of false positives \citep[e.g.][]{cw09, mathur12, barclay15}.  In recent years, some progress has been made by applying asteroseismic techniques to suitable evolved stars with transiting planet candidates \citep[e.g.][]{quinn15, grunblatt17, TOI197, chontos19}, but it seems likely that the problems inherent to detecting transiting planets orbiting evolved stars will continue to confound observers through the coming years.

To study the occurrence rates of planets orbiting evolved stars, other methods are needed.  A number of radial velocity surveys have been targeting evolved stars for almost 20 years, with the main scientific goal being to understand the properties of planetary systems orbiting stars more massive than our Sun. Those surveys began in an attempt to circumvent the challenges inherent to the radial velocity detection of planets orbiting massive stars. 

The technical requirements imposed by Doppler exoplanetary detection mean that the most favourable main sequence target stars lie in a narrow range of masses centred on 1\,\Msun.  Stars in this Sun-like mass range are cool enough and rotate slowly enough to present an abundance of narrow spectral absorption lines for accurate velocity determination.  In contrast, however, more massive stars on the main sequence are too hot and rotate too rapidly for this technique to work.  Main sequence stars of higher mass have few usable absorption lines (due to their high temperatures), and also tend to be fast rotators \citep[$v \sin i$ > 50 km/s;][]{galland05} - which causes what spectral lines they do have to be sufficiently broad as to render them useless for the detection of planet-mass objects.  In addition, the shorter main-sequence lifetimes of higher-mass stars means that they will preferentially be observed at younger ages. Furthermore, stars earlier than a spectral class of around F7 also have much shallower convection zones than later-type stars, and so do not experience the magnetic braking which slows the rotation of those later-type (lower-mass) stars.  

Whilst massive main-sequence stars are poor choices for radial velocity observations, their evolved siblings present a far better target.  In particular, subgiants and low-luminosity giants are ideal radial velocity targets because their surface gravities remain high enough (log $g\gtsimeq$3) to avoid the large-amplitude pulsations common in red giants \citep{hekker08}, whilst still rotating slowly, and being cool enough to have the abundant, narrow spectral lines that facilitate radial velocity observations. 

To learn more about the occurrence and properties of planets around more massive stars, several teams have been surveying so-called ``retired A stars,'' with a combined total of $\sim$1000 targets and ten to fifteen years of observations \citep[e.g.][]{sato05, johnson06, jones11, reffert15}. These surveys have borne fruit, with more than 100 planets being found to date \citep[e.g.][]{johnson07, johnson11, jones11, sato12, 121056, n15, luhn19}.  As a result, we are now beginning to understand the relationship between stellar mass and the abundance of giant planets, with strong indications that giant planets are more efficiently formed around more massive stars \citep[e.g.][]{bowler10, maldonado13, jones16, witt17a}.  

The Pan-Pacific Planet Search \citep[PPPS;][]{47205} is an international collaboration between Australia, China, and the US, with the aim of attacking this critical problem by obtaining precision radial-velocity measurements of bright Southern hemisphere, evolved intermediate-mass stars.  The mean properties of the PPPS sample, as fully detailed in \citet{ppps5}, are 1.31$^{+0.28}_{-0.25}$\,\Msun, log $g=3.09\pm$0.26 dex, [Fe/H]$=-0.03\pm$0.16 dex, and $T_{eff}=4812\pm$166\,K.  The PPPS operated on the Anglo-Australian Telescope from 2009-2014, contributing to the discovery of 15 planets orbiting evolved stars \citep{47205, 121056, sato13, ppps3, ppps4}.  Unfortunately, due to shifting priorities in the Australian telescope time assignment process, this program and the eighteen-year Anglo-Australian Planet Search \citep{tinney01} were prematurely terminated in 2014, and many PPPS targets were left with inadequate sampling to confirm or refute emerging candidate signals.  The PPPS had 37 targets in common with the EXPRESS survey of southern evolved stars \citep{jones11, jones14}, and in recent years we have jointly published several planet discoveries where our combined data sets confirmed the signals seen in the data from one or other of those surveys \citep{jones16, jones17, witt17a}, results that have included the most eccentric planet known to orbit an evolved star \citep{76920}.   

This paper is the final instalment of the PPPS series.  We release all the final radial velocity measurements in Section 2, and in Section 3 describe a handful of potential candidates that require further observations to confirm.  In Section 4, we perform an analysis of the detection limits from this survey and derive an estimate of the occurrence rate of giant planets orbiting evolved stars, before drawing our conclusions in Section 5.

%-----------------------------------------------------------------------
\section{Observational Data}

%\subsection{Radial Velocity Observations}

We observed the PPPS target stars using the UCLES spectrograph \citep{diego:90} on the 3.9m Anglo-Australian Telescope from 2009 February until 2015 January.  UCLES achieved a resolution of 45,000 with a 1 arcsec slit, and we aimed to achieve a signal-to-noise ratio (S/N) of 100 at 5500 \AA\ per spectral pixel at each epoch, resulting in exposure times ranging from 100-1200s.  An iodine absorption cell provided wavelength calibration from 5000 to 6200\AA.  The spectrograph point-spread function and wavelength calibration are derived from the iodine absorption lines embedded on every pixel of the spectrum by the cell \citep{val:95,BuMaWi96}.  The result is a precision Doppler velocity estimate for each epoch, along with an internal uncertainty estimate, which includes the effects of photon counting uncertainties, residual errors in the spectrograph PSF model, and variation in the underlying spectrum between the iodine-free template and epoch spectra observed through the iodine cell.  The photon-weighted mid-time of each exposure is determined by an exposure meter.  All velocities are measured relative to the zero point defined by the template observation.  The iodine-free template spectrum for each star was obtained with the 0.75\arcsec\ slit for a resolution of 60,000 with $S/N\sim$150-300 per pixel.  Table~\ref{tab:allvels} gives the complete set of final radial velocities from 105 PPPS targets.  Table~\ref{tab:dispositions} summarises the final dispositions of all PPPS targets, e.g. published companion, candidate, or double-lined binary.

%-----------------------------------------------------------------------
\section{Candidate Signals}

\begin{figure*}
	\includegraphics[width=\columnwidth]{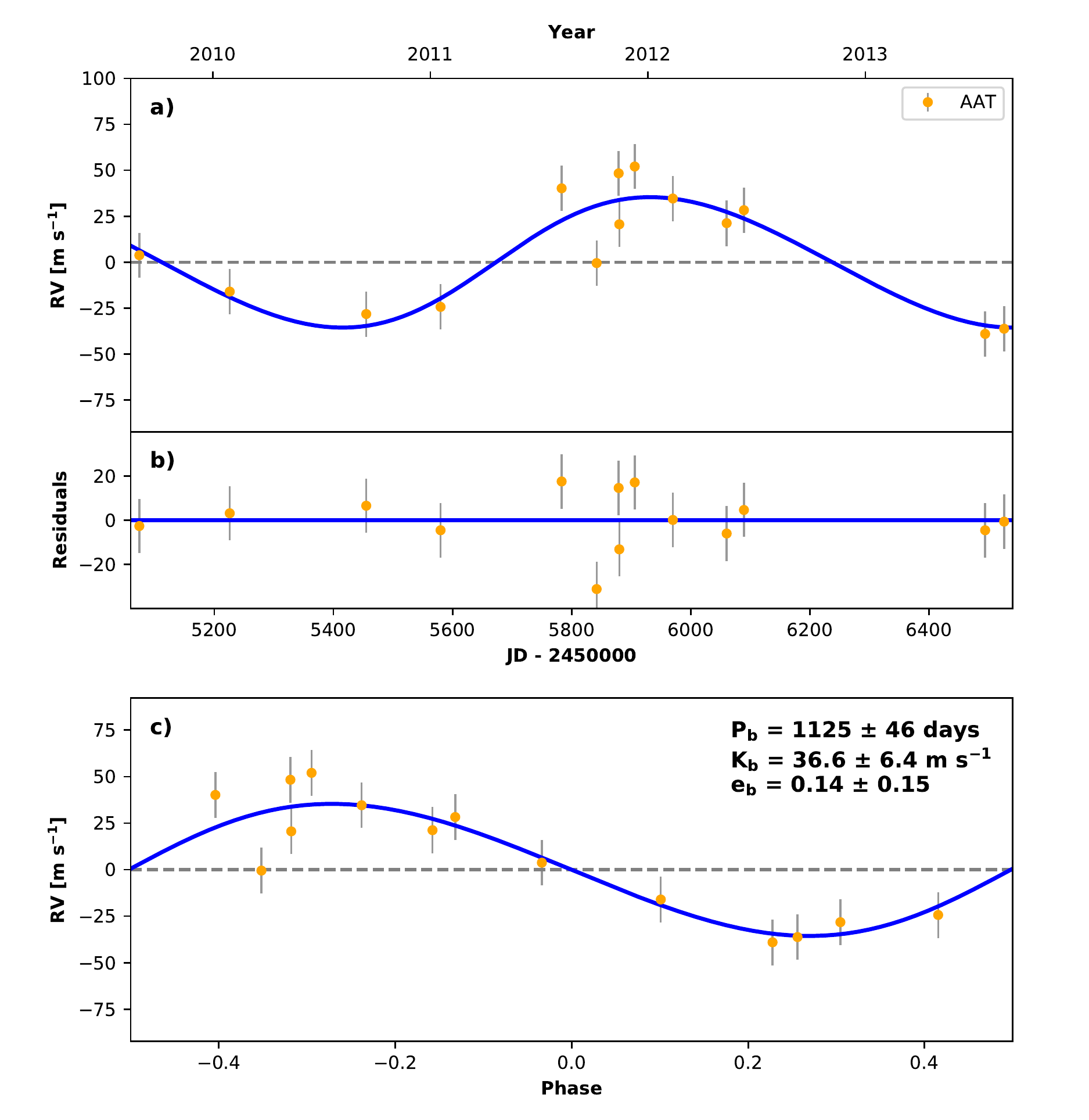}
	\includegraphics[width=\columnwidth]{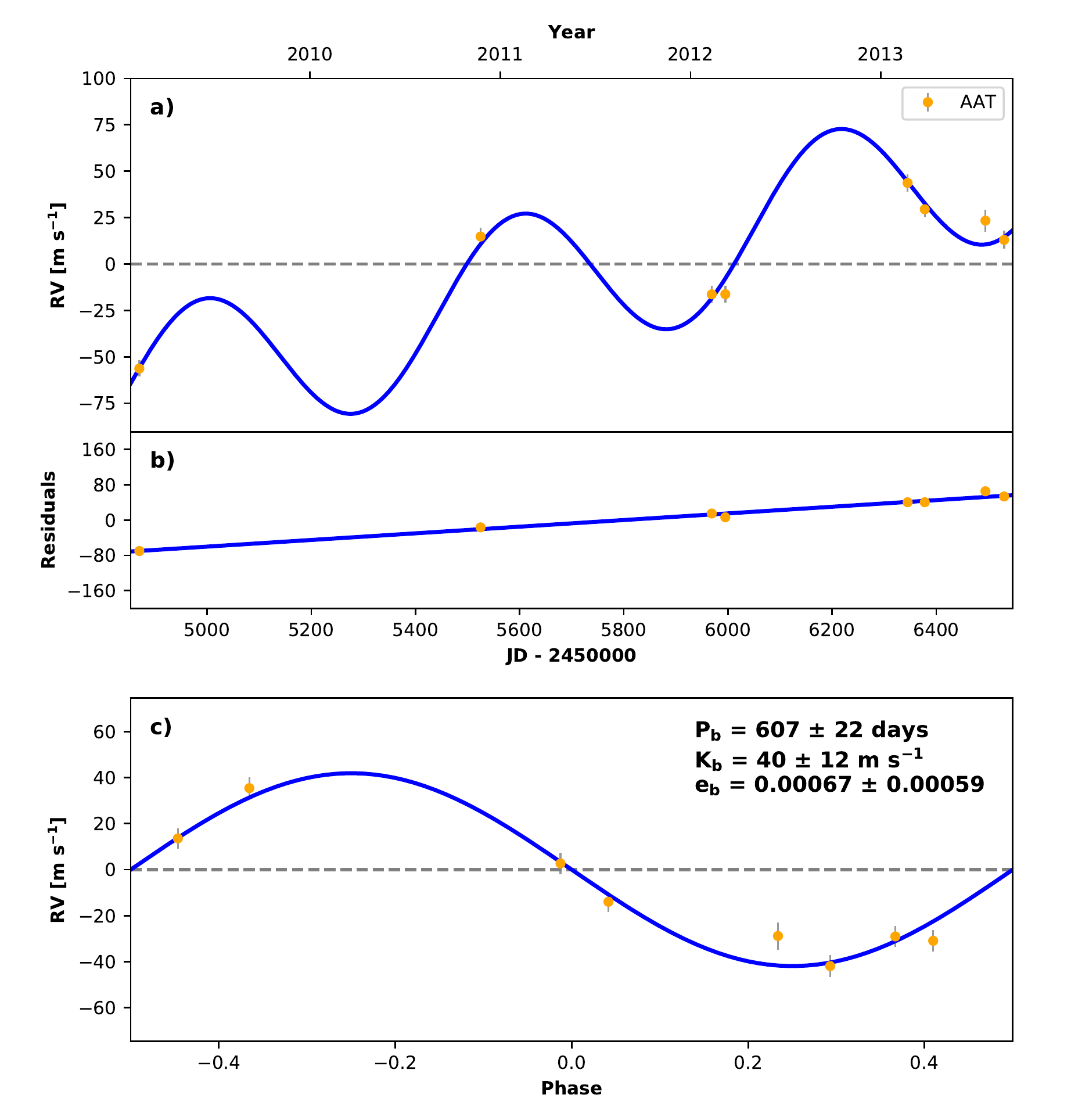}\\
	\includegraphics[width=\columnwidth]{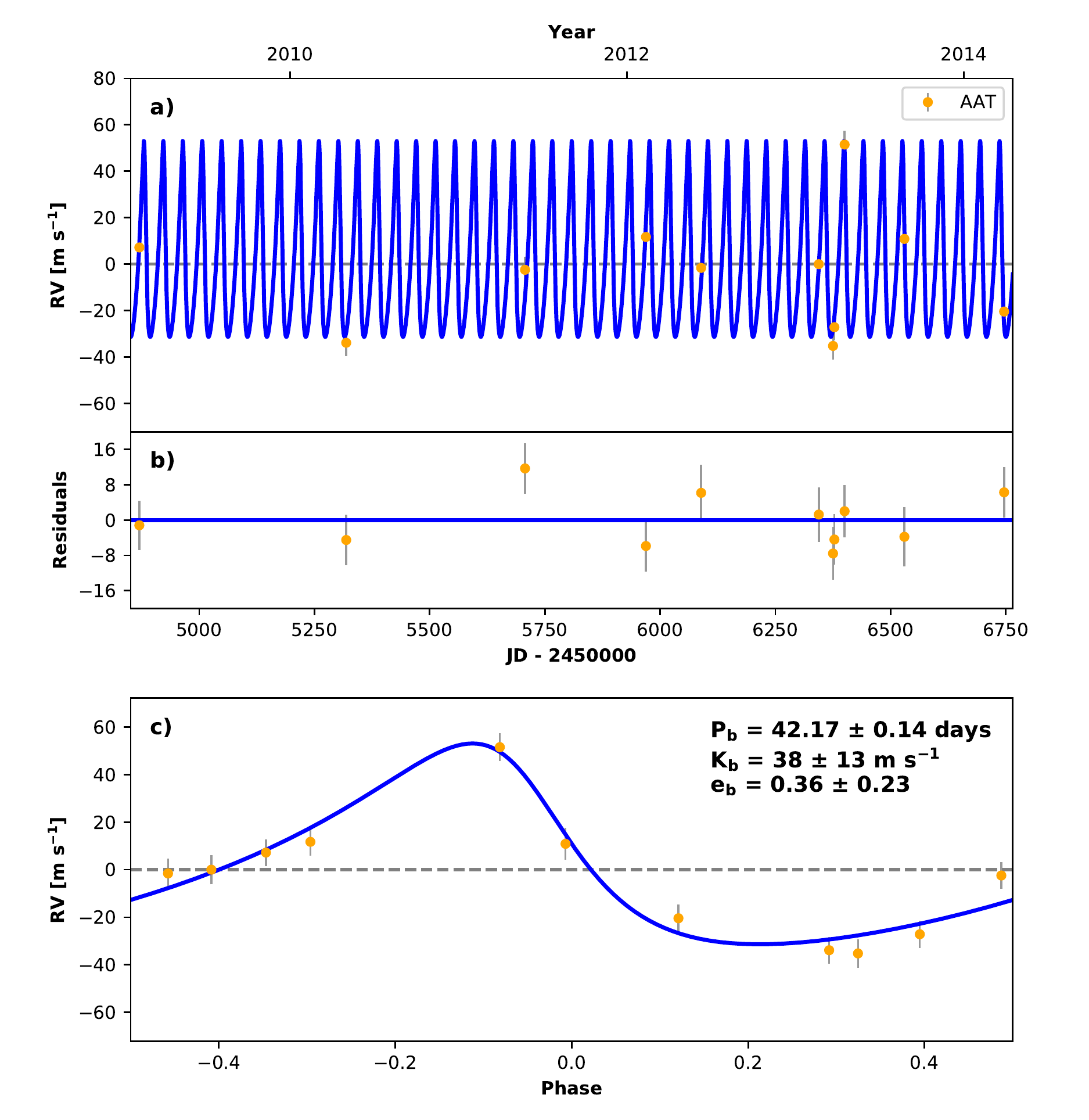}
	\includegraphics[width=\columnwidth]{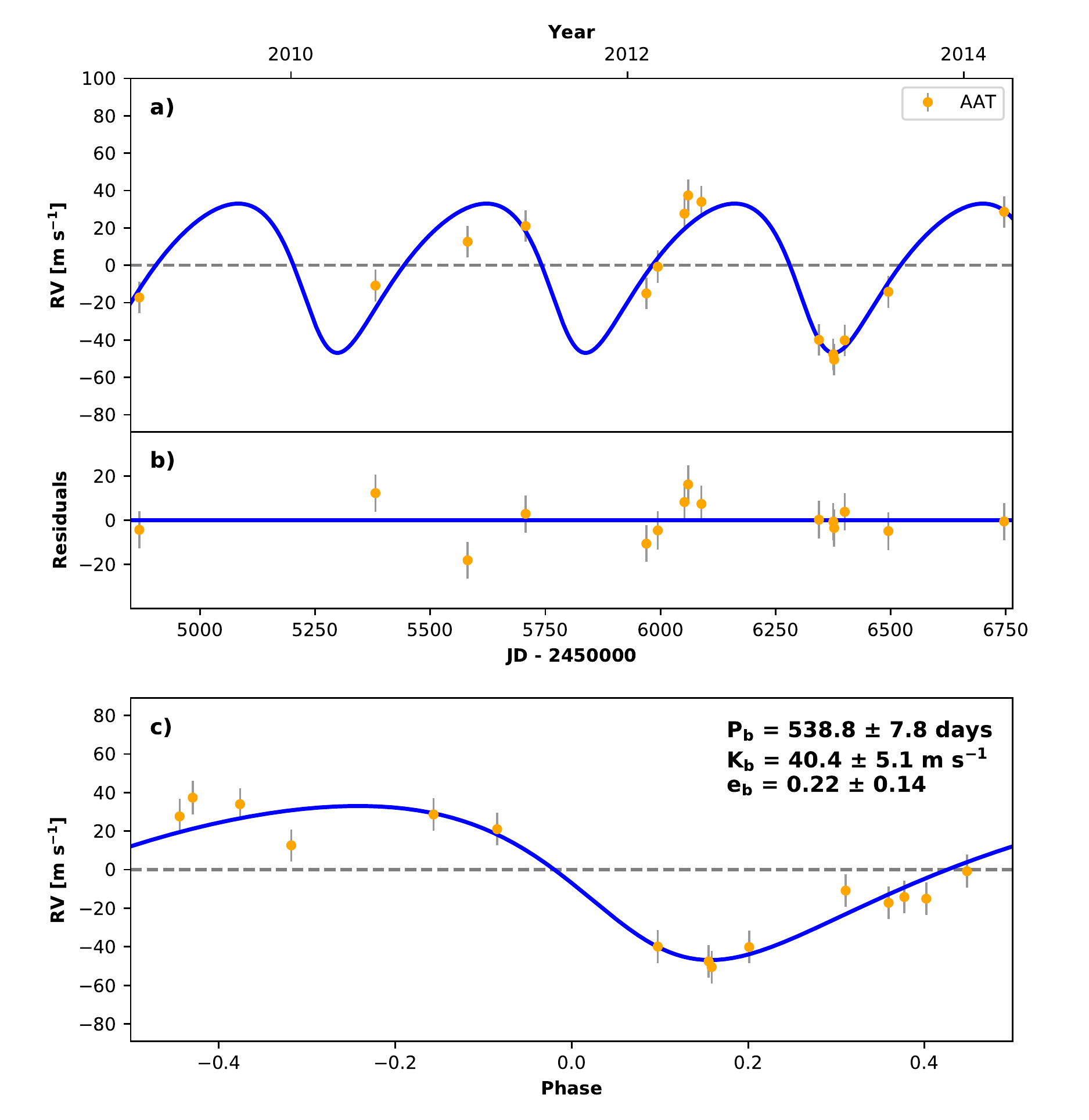}\\
    \caption{Data and model fits for candidates from the PPPS.  Fits shown are tentative and require further observations to be confirmed.  Clockwise from top left: HD\,6037, HD\,13652 (RV trend included), HD\,114899, HD\,126105.  For each candidate, we show the time series and phase-folded fits.}
    \label{fig:fits1}
\end{figure*}

\begin{figure*}
	\includegraphics[width=\columnwidth]{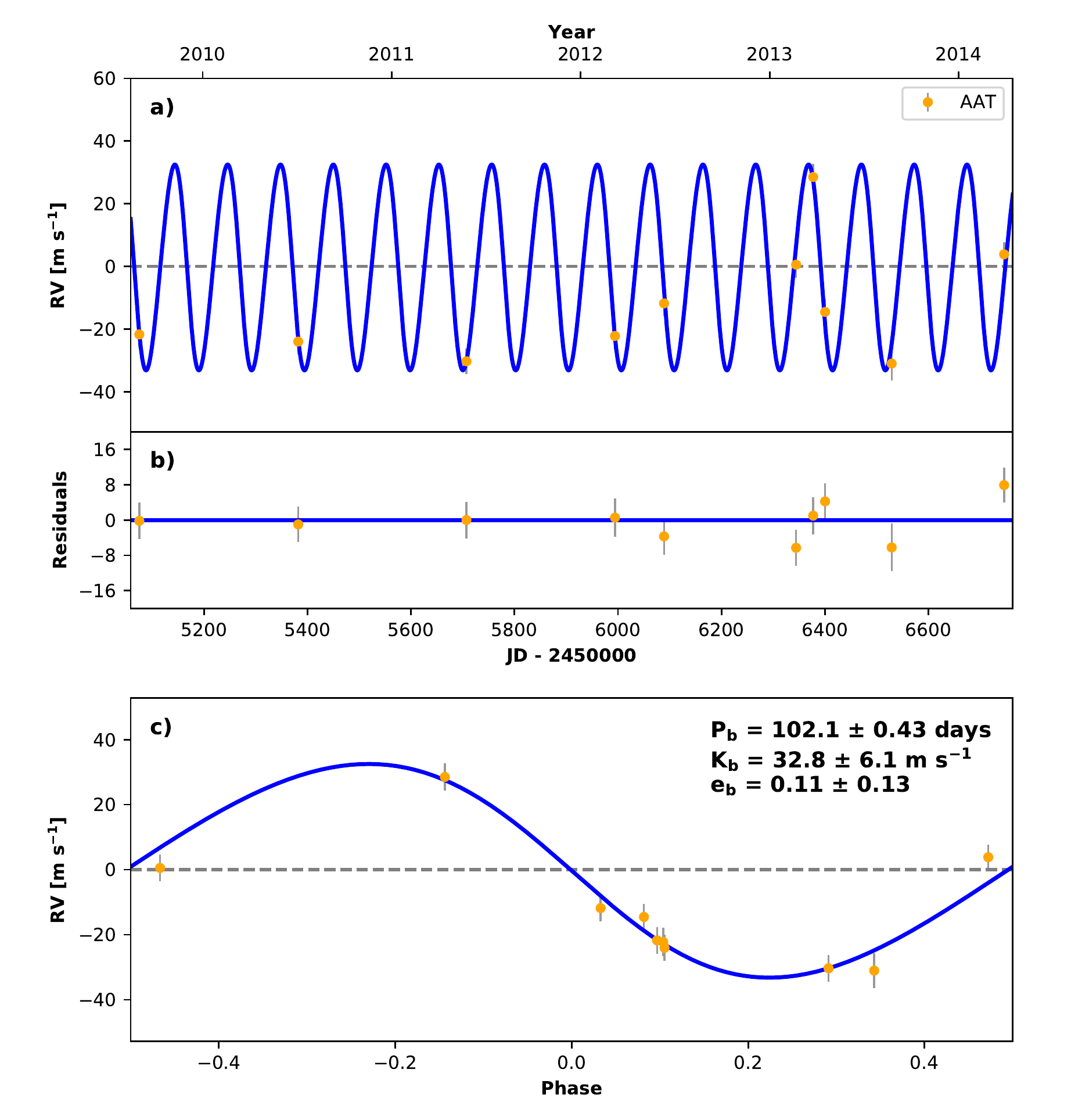}
	\includegraphics[width=\columnwidth]{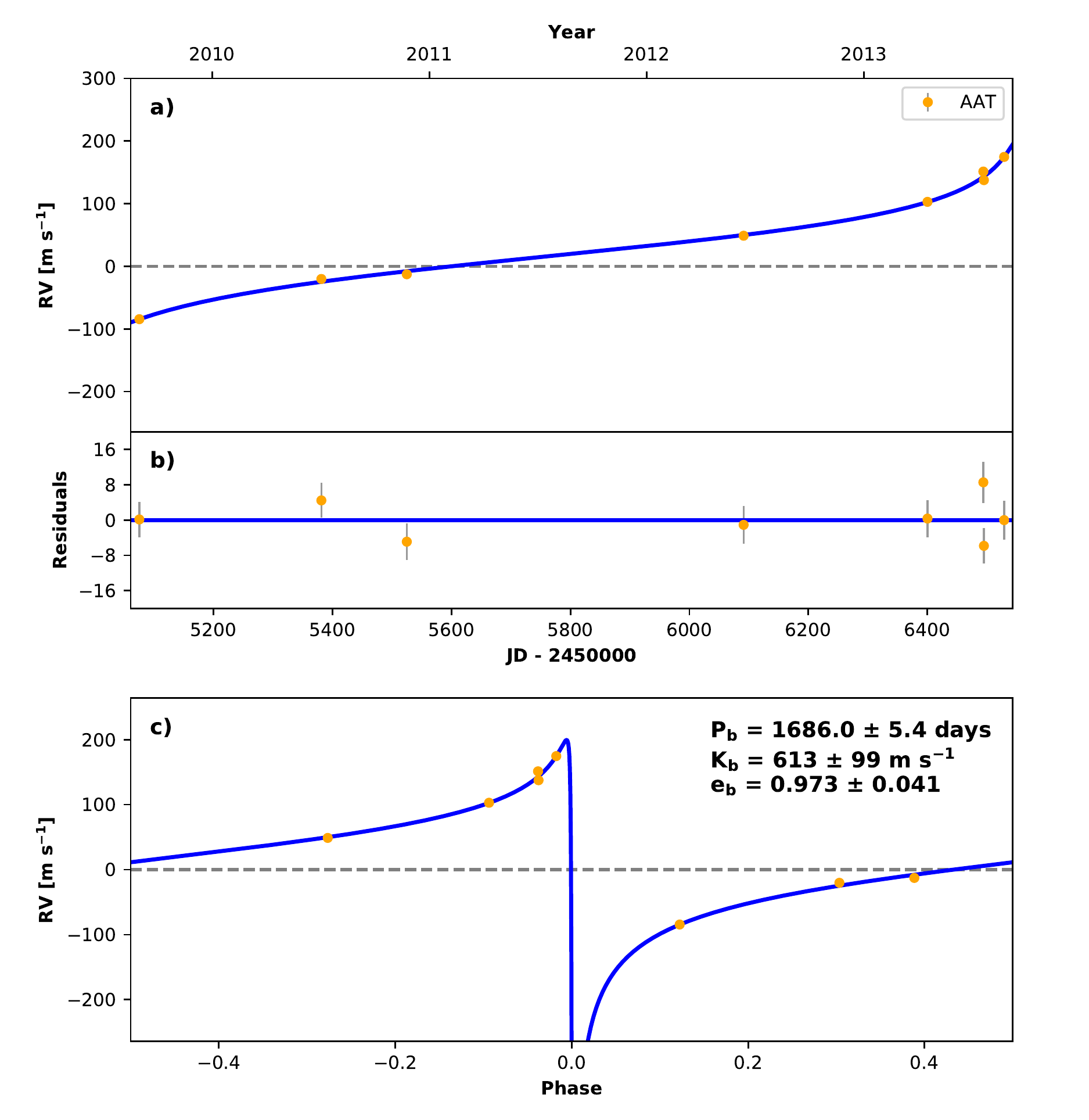}\\
	\includegraphics[width=\columnwidth]{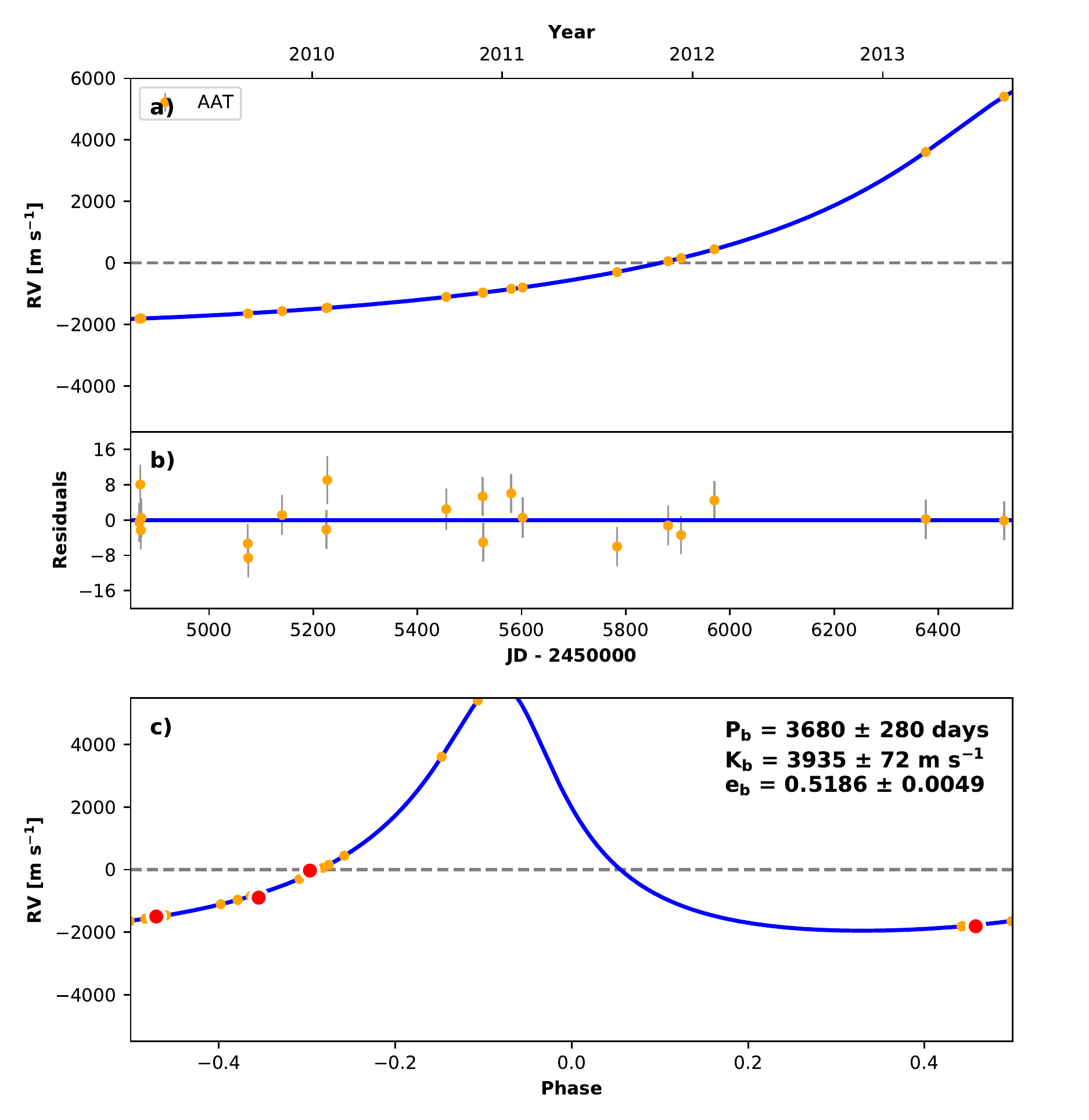}
	\includegraphics[width=\columnwidth]{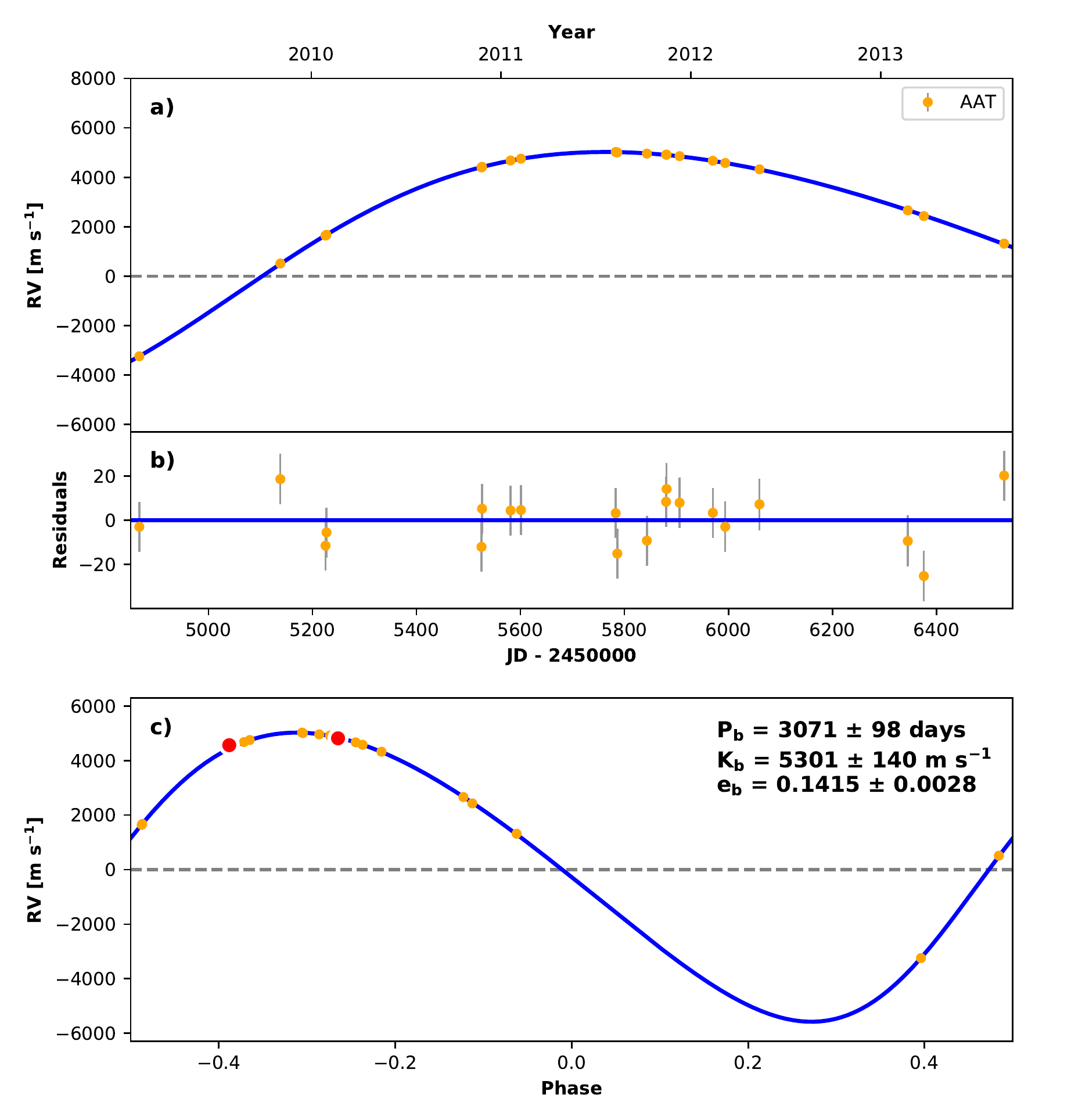}\\
    \caption{Data and model fits for candidates from the PPPS.  Fits shown are tentative and require further observations to be confirmed.  Clockwise from top left: HD\,159743, HD\,205577, HD\,37763, HD\,43429.  For each candidate, we show the time series and phase-folded fits. }
    \label{fig:fits2}
\end{figure*}

\begin{figure*}
	\includegraphics[width=\columnwidth]{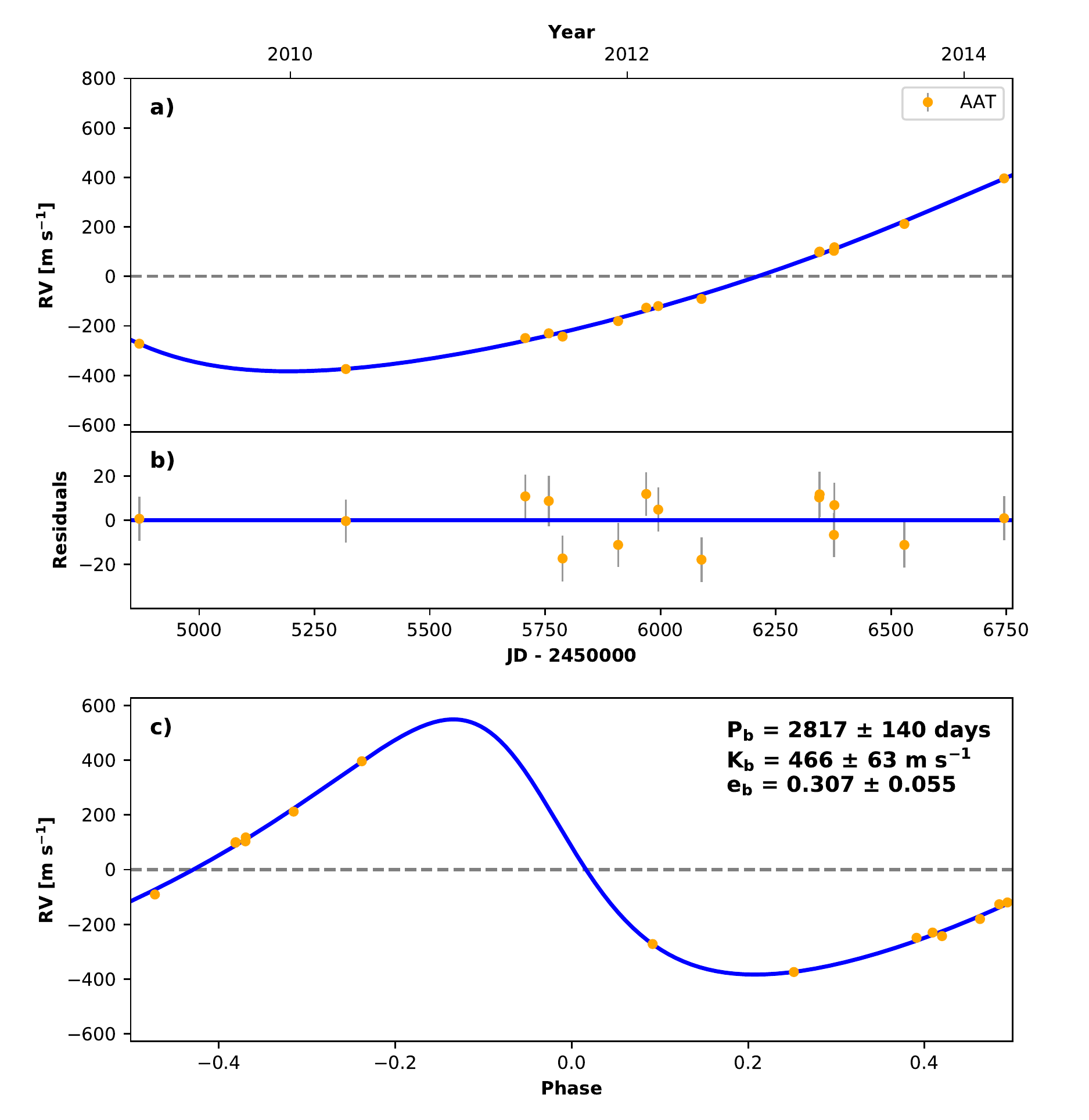}
	\includegraphics[width=\columnwidth]{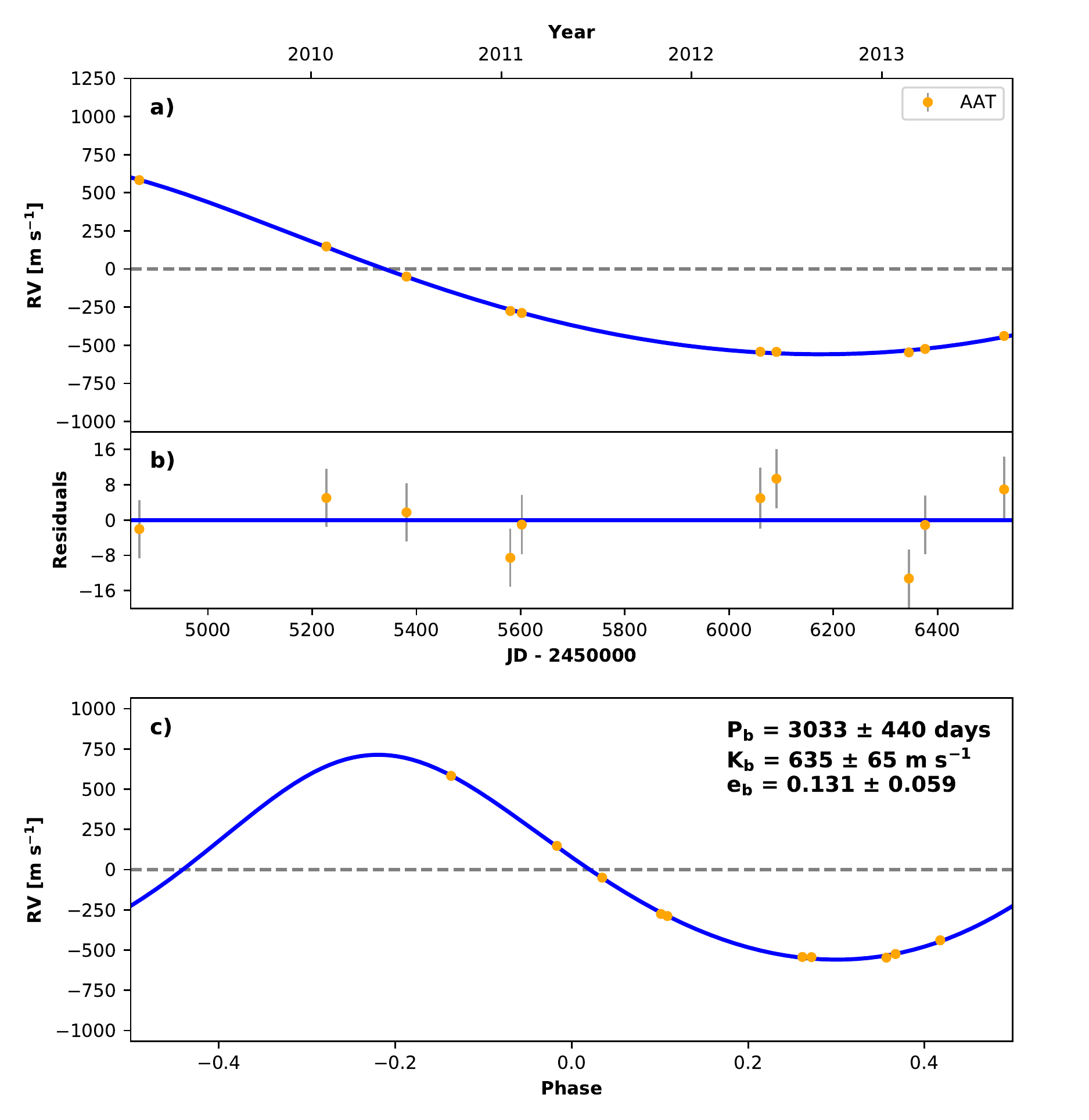}\\
	\includegraphics[width=\columnwidth]{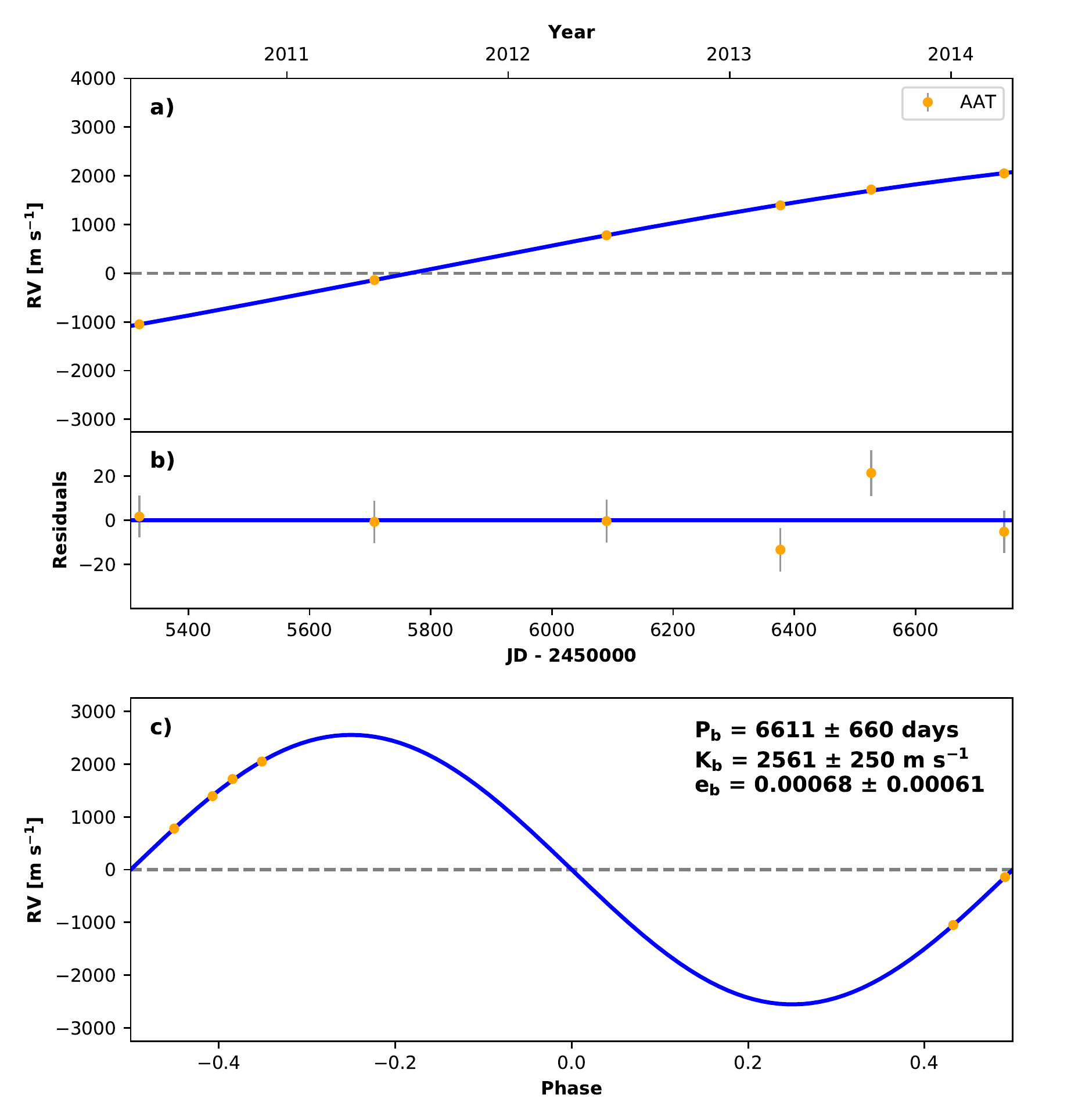}
	\includegraphics[width=\columnwidth]{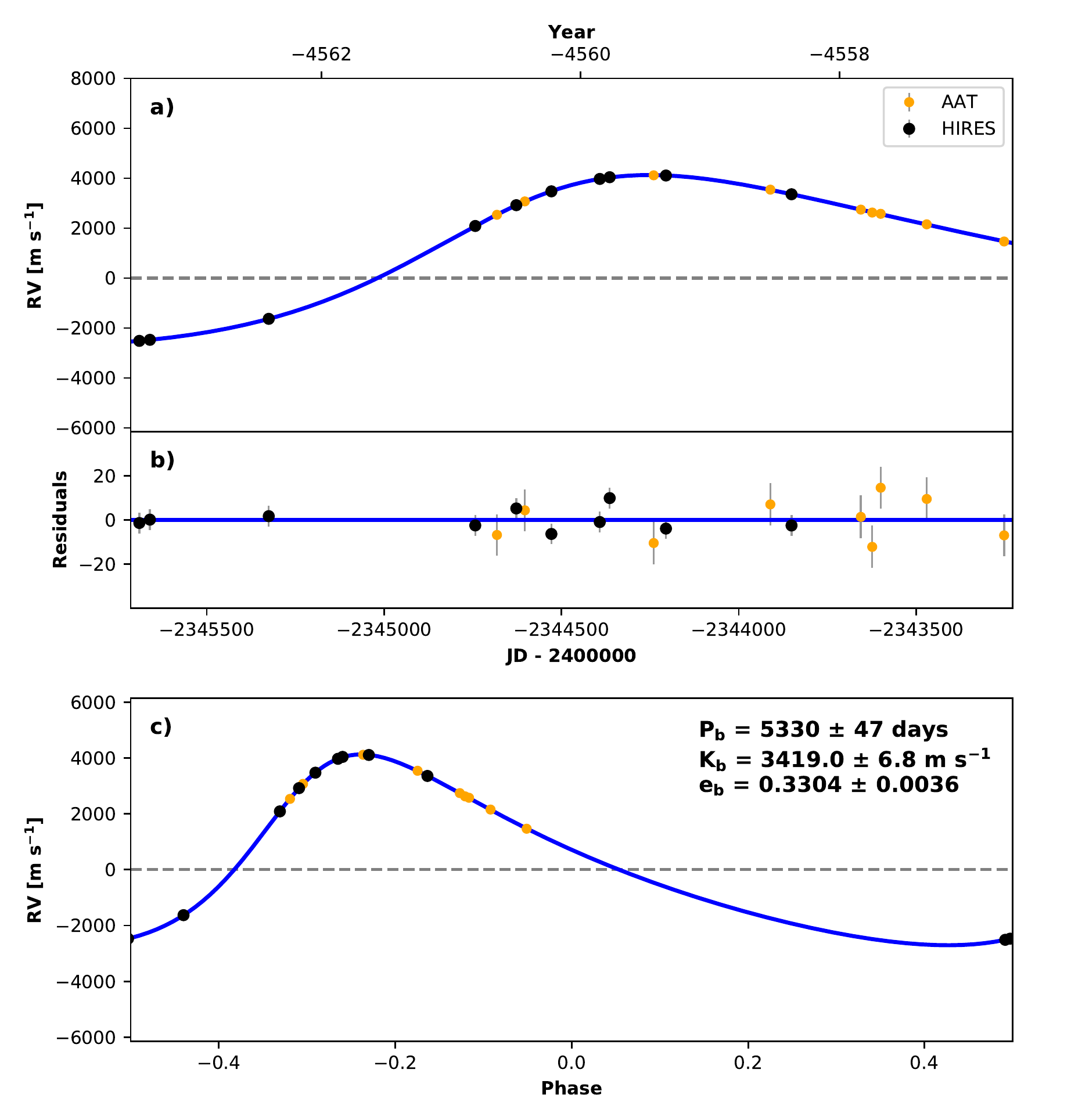}\\
    \caption{Data and model fits for candidates from the PPPS.  Fits shown are tentative and require further observations to be confirmed.  Clockwise from top left: HD\,115066, HD\,121156, HD\,142132, HD\,145428.  For each candidate, we show the time series and phase-folded fits. }
    \label{fig:fits3}
\end{figure*}

Whilst all of the secure planet detections from this survey have been published, the truncated temporal nature of our dataset makes it 
%it is 
inevitable that some stars will exhibit RV variations suggestive of substellar companions that still require the acquisition of additional data to either confirm or refute.
%, requiring more data to confirm or refute.  
Since the main PPPS survey has been concluded, in the interest of completeness, we now describe 12 potential candidates that may warrant further follow-up.  These candidates fall into two broad categories: those for which a tentative orbital period can be obtained, and unconstrained long-period signals.  They were identified by examining those stars which had (1) at least 8 RV epochs (to enable a nontrivial Keplerian fit attempt), and (2) RMS exceeding 15\,\ms\ (about three times the typical jitter for these stars).  We performed initial searches on those targets using a genetic algorithm to fit a single Keplerian orbit.  If this resulted in a mass detection of at least $3\sigma$ with no large phase gaps, then we performed more detailed fits including a full MCMC parameter determination.  Table~\ref{tab:fantasyfits}, divided into substellar and stellar-mass candidates, gives the best-fit parameters as derived from \textit{RadVel} \citep{fulton18}.  We emphasize that at this time we cannot claim these objects to be confirmed companions, and we show these example fits merely to guide future follow-up efforts.  Figures~\ref{fig:fits1}--\ref{fig:fits3} show the data and the best fits for those where a plausible unique orbital solution could be obtained.  Candidate minimum masses ($m$ sin $i$) were derived from the host-star masses as presented in \citet{ppps5}, which presented complete spectroscopic stellar parameters for the PPPS sample. 

% systemic bootstrap results commented out here.
% updated with MY radvel results.
\begin{table*}
	\centering
	\caption{Orbital solutions for candidate companions.}
	\label{tab:fantasyfits}
	\begin{tabular}{llllllll} 
		\hline
		       & Period & Eccentricity & $\omega$ & $T_c$ & $K$ & m sin $i$ & $a$ \\
		Host & days &   & degrees & BJD-2400000 & \ms & \Mjup & au \\
		\hline
		%HD 6037 & 1111$\pm$74 & 0.1$\pm$0.3 & 341$\pm$67 & 49236$\pm$384 & 37.0$\pm$9.5 & 2.4$\pm$0.4 & 2.37$\pm$0.11 \\
		HD 6037 & 1125$^{+47}_{-44}$ & 0.1$^{+0.2}_{-0.1}$ & 354$^{+229}_{-206}$ & 55112$^{+71}_{-66}$ & 36.6$^{+6.8}_{-6.1}$ & 2.4$\pm$0.5 & 2.39$\pm$0.07 \\
		%HD 13652 & 626$\pm$54 & 0.86$\pm$0.13 & 341$\pm$48 & ??? ma 329 45 &  & 7.0$\pm$2.7 & 1.54$\pm$0.09 \\ % needs slope too. High e fit
		%HD 13652 & 569$\pm$58 & 0.0 (fixed) & 0.0 (fixed) & ??? ma 231 50 & 42$\pm$12 & 2.0$\pm$0.7 & 1.45$\pm$0.10 \\ % needs slope too. adopted circular fit.
		HD 13652 & 607$\pm$22 & 0.0 (fixed) & 0.0 (fixed) & 54533$^{+62}_{-63}$ & 40$^{+11}_{-12}$ & 1.9$\pm$0.7 & 1.51$\pm$0.05 \\ 
		%HD 114899 & 42.19$\pm$0.07 & 0.36$\pm$0.27 & 49$\pm$75 & 49988$\pm$9 & 41.1$\pm$136 & 0.9$\pm$1.0 & 0.2722$\pm$0.0003 \\
		HD 114899 & 42.17$\pm$0.14 & 0.36$^{+0.25}_{-0.2}$ & 52$^{+74}_{-57}$ & 55097.7$^{+5.6}_{-4.4}$ & 38$^{+10}_{-15}$ & 0.8$\pm$0.2 & 0.272$\pm$0.003 \\
		%HD 126105 & 539$\pm$11 & 0.29$\pm$0.20 & 131$\pm$34 & 49868$\pm$158 & 38$\pm$7 & 1.55$\pm$0.35 & 1.33$\pm$0.02 \\
		HD 126105 & 538.8$^{+7.6}_{-7.9}$ & 0.22$^{+0.15}_{-0.13}$ & 129$^{+37}_{-44}$ & 55204$^{+26}_{-24}$ & 40.4$^{+5.4}_{-4.9}$ & 1.55$\pm$0.35 & 1.33$\pm$0.02 \\
		%HD 159743 & 102.0$\pm$0.4 & 0.12$\pm$0.33 & 113$\pm$98 & 49971$\pm$26 & 32$\pm$78 & 0.93$\pm$0.77 & 0.484$\pm$0.002 \\  overlap with Matias
		HD 159743 & 102.1$^{+0.47}_{-0.40}$ & 0.12$^{+0.18}_{-0.08}$ & 29$^{+178}_{-281}$ & 55063.7$^{+5.0}_{-6.7}$ & 32.8$^{+5.8}_{-6.4}$ & 0.96$\pm$0.19 & 0.484$\pm$0.007 \\  
		HD 205577 & 1685.98$^{+11.0}_{-0.09}$ & 0.972$^{+0.08}_{-0.002}$ & 127$^{+16}_{-80}$ & 48155$^{+57}_{-51}$ & 613$^{+47}_{-150}$ & 9.3$\pm$2.3 & 2.87$\pm$0.05 \\
		\hline
		%HD 37763 & 3464$\pm$598 & 0.514$\pm$0.017 & 16$\pm$8 & 49721$\pm$847 & 3888$\pm$274 & 300$\pm$52 & 4.9$\pm$0.7 \\
		HD 37763 & 3680$^{+330}_{-240}$ & 0.52$\pm$0.01 & 13$\pm$3 & 53241$^{+200}_{-270}$ & 3935$^{+82}_{-62}$ & 262$\pm$20 & 5.1$\pm$0.3 \\
		%HD 43429 & 3350$\pm$381 & 0.140$\pm$0.005 & 261$\pm$16 & 48297$\pm$630 & 5677$\pm$562 & 648$\pm$110 & 5.5$\pm$0.5 \\ 
		HD 43429 & 3071$^{+96}_{-100}$ & 0.142$\pm$0.003 & 248$\pm$5 & 53651$^{+81}_{-77}$ & 5301$^{+140}_{-150}$ & 456$\pm$29 & 5.0$\pm$0.1 \\
		%HD 51268 & 5960$\pm$1547 & 0.21$\pm$0.18 & 121$\pm$38 & ????  & 890$\pm$260 & 78$\pm$29 & 6.6$\pm$1.2 \\ curvature nonzero ecc but radvel won't plot.
		%HD 51268 & 6668$^{+240}_{-230}$ & 0.21 (fixed) & 121 (fixed) & 50591$\pm$150 & 920$^{+300}_{-260}$ & 84$\pm$31 & 7.1$\pm$0.3 \\ pain in arse, demote to quadratic only...
		%HD 84070 & 2158$\pm$49 & 0.90$\pm$0.25 & 33$\pm$3 & 48712$\pm$90 & 23642$\pm$3236 & 1242$\pm$134 & 4.5$\pm$0.1 \\  DO NOT SHOW THIS IT IS BULLSHIT
		%HD 115066 & 2749$\pm$185 & 0.31$\pm$0.15 & 55$\pm$11 & 48637$\pm$937 & 460$\pm$161 & 34$\pm$14 & 4.1$\pm$0.3 \\ % overlap with Matias
		HD 115066 & 2817$\pm$140 & 0.31$^{+0.06}_{-0.05}$ & 53$^{+5}_{-6}$ & 54610$^{+45}_{-53}$ & 466$^{+78}_{-47}$ & 35$\pm$7 & 4.2$\pm$0.2 \\ 
		%HD 121156 & 2383$\pm$655 & 0.25$\pm$0.15 & 333$\pm$12 & 49967$\pm$1230 & 573$\pm$220 & 45$\pm$21 & 3.9$\pm$0.7 \\ % overlap with Matias
		HD 121156 & 3033$^{+470}_{-420}$ & 0.13$^{+0.07}_{-0.05}$ & 345$^{+28}_{-14}$ & 55280$^{+37}_{-67}$ & 635$^{+82}_{-48}$ & 54$\pm$11 & 4.6$\pm$0.4 \\  
		%HD 124087 & 5922$\pm$914 & 0.0 (fixed) & 0.0 (fixed) & ??? ma 310$\pm$41 & 1659$\pm$200 & 166$\pm$33 & 7.1$\pm$0.7 \\ %curvature quad fit ONLY do not show.
		%HD 142132 & 6589$\pm$858 & 0.0 (fixed) & 0.0 (fixed) & 44237$\pm$1614 & 2545$\pm$314 & 317$\pm$59 & 8.0$\pm$0.8 \\ % overlap with Matias
		HD 142132 & 6611$^{+720}_{-600}$ & 0.0 (fixed) & 0.0 (fixed) & 59063$^{+300}_{-250}$ & 2561$^{+270}_{-220}$ & 277$\pm$47 & 8.0$\pm$0.5 \\ %curvature only, fix e=0
		%HD 145428 & 5335$\pm$96 & 0.331$\pm$0.008 & 309.1$\pm$0.7 & 45269$\pm$2053 & 3420$\pm$10 & 397$\pm$3 & 7.14$\pm$0.09 \\ %Luhn+2018 with keck
		HD 145428 & 5335$\pm$96 & 0.331$\pm$0.008 & 309.1$\pm$0.7 & 45269$\pm$2053 & 3420$\pm$10 & 397$\pm$3 & 7.14$\pm$0.09 \\ 
		\hline
	\end{tabular}
		\\
\end{table*}

% 124087 - quad fit as good as any Keplerian.
% 84070 - not enough to do shit
% 51268 - e=0.22 gets 12.5 m/s rms vs quad fit at 21.8 m/s.

% Radvel bug: If gamma >1000 it just refuses to plot.

%\clearpage

%\clearpage

For those stars with potential stellar-mass companions, we checked the \textit{Gaia} DR2 results for astrometric or RV signatures of hidden massive bodies.  The results of that search are summarised in Table~\ref{tab:bigassthings}.  The lower section of Table~\ref{tab:bigassthings} gives the \textit{Gaia} DR2 notes for the 12 stars in the PPPS sample which show large (\kms) RV variations indicative of stellar-mass companions, but for which we have too few observations to attempt an orbital solution.  For nearby stars, high-contrast imaging can resolve the influencing body, yielding better constraints on the system parameters \citep[e.g.][]{crepp12, rodigas16, kane19}, even when the object is not seen \citep[e.g.][]{whitedwarf, hirsch19}.  However, with our targets generally falling at distances of 150-300\,pc, we do not expect that any stellar companions could typically be resolved in \textit{Gaia} imaging, though we note that HD\,110238 has a \textit{Gaia} detected companion with common proper motion which is the likely cause of the observed large-amplitude radial velocity variation in our PPPS data.  For the very bright stars considered here ($G<8$), the expected \textit{Gaia} RV precision is typically better than about 0.4\,\kms\ \citep{katz19}.  Stars exhibiting significantly higher uncertainties in their measured absolute radial velocity may indicate binarity.  We flag here those stars with RV errors more than 3$\sigma$ too large.  We have begun additional monitoring of the candidates in Table~\ref{tab:fantasyfits} as a ``PPPS Legacy" program with the \textsc{Minerva}-Australis dedicated telescope array \citep{witt18,addison19}.  For these bright stars and large candidate signals, \textsc{Minerva}-Australis is easily able to obtain new precise RV measurements \citep[e.g.][]{nielsen19, vanderburg19} over the coming years to clarify the nature of these objects.

% 84070, 124087 are not fit as Keplerians
% 51268 needs an eccentric Keplerian but radvel refuses to plot, gamma too large. I cannot kludge a workaround bugger it, move that to 'unconstrained shit' in lower section of table.

\begin{table*}
	\centering
	\caption{\textit{Gaia} DR2 notes on potential stellar-mass companions.}
	\label{tab:bigassthings}
	\begin{tabular}{lll} 
		\hline
		Star &  m sin $i$ & Notes \\
		  & \Msun\ &  \\
		\hline
		HD\,37763 & 0.29  & No excess astrometric noise \\
		HD\,43429 & 0.62  & 54.7$\sigma$ excess astrometric noise \\
		HD\,142132 & 0.30 & No excess astrometric noise \\
		HD\,145428 & 0.38 & No excess astrometric noise \\
		\hline
		HD\,5676 &  & No excess astrometric noise \\
		HD\,11653 &  & No excess astrometric noise \\
		HD\,14791 &  & 55.5$\sigma$ excess astrometric noise \\
		HD\,51268 &  &  No excess astrometric noise  \\
		HD\,84070 &  & 349.6$\sigma$ excess astrometric noise \\
		HD\,104819 &  & \textit{Gaia} RV error 5$\sigma$ too large. 46.7$\sigma$ excess astrometric noise \\
		HD\,110238 &  & \textit{Gaia} RV error 8.3$\sigma$ too large. CPM companion at $\Delta\,G=8.9$ \\
		HD\,124087 &  & No excess astrometric noise \\
		HD\,166309 &  & No excess astrometric noise \\
		HD\,181809 &  & \textit{Gaia} RV error 7.9$\sigma$ too large. \\
		HD\,204057 &  & \textit{Gaia} RV error 11$\sigma$ too large. \\
		HD\,222768 &  & \textit{Gaia} RV error 4.1$\sigma$ too large. \\
		\hline
	\end{tabular}
\end{table*}

%The UCLES spectrograph has been decommissioned, with the new Veloce high-resolution stabilised spectrograph replacing it at the AAT \citep{veloce}.

%-----------------------------------------------------------------------
\section{Detection Limits}

Often overshadowed by discoveries, the use of observational data to determine what was \textit{not} found is of at least equal importance to the advancement of our understanding of exoplanetary populations.  Previous work in this area has had the luxury of large amounts of data, derived from legacy RV surveys where it was eminently reasonable to impose minimum thresholds for the number of observations.  Traditional injection-recovery tests have usually set a minimum of $N\sim$30 RV data points to derive reliable detection limits \citep[e.g.][]{witt06,cumming08,jupiters}.  Similar efforts to derive detectabilities and occurrence rates from space-based photometry, such as that obtained by the \textit{Kepler} mission, can make use of many thousands of observations to compute a sensitivity function \citep[e.g.][]{jessie12, jessie16, coughlin16, zink19}.  However, the field of radial velocity exoplanet detection is littered with the desiccated husks of surveys cut short before large quantities of data were obtained.  In this section, we describe our efforts to glean useful constraints on the planets that can be excluded by our PPPS data.  The median number of RV epochs in the PPPS data considered here is 8.  Traditional periodogram approaches used to recover injected signals simply fail in this sparse regime.

\subsection{Techniques}

\citet{meunier12} compared the performance of several detection limit methods on RV data sets from ten stars with a variety of properties.  Of interest for the present work are two methods which do not rely on the use of periodograms.  First is the root-mean-square (RMS) method, based on the principles outlined in \citet{galland05} and reprised briefly in \citet{meunier12}.  For 1000 trial phases of a simulated planetary RV signal with a given period and mass (i.e. an RV amplitude, $K$), we ask whether the RMS of the simulated RV data (that of a planetary orbit sampled at the timestamps of the real data) is greater than the RMS of the original data.  If all 1000 such realisations give an RMS higher than the real data, then we say that planetary signal is excluded by the data at 99.9\% confidence.  Second is the F-test method, which is at its core an injection-recovery approach, except that the criterion for determining whether a signal is detectable is the F-test rather than a periodogram.  We add a simulated planetary RV signal to the data, then perform an F-test to ask whether the two data sets (original and with added planetary signal) are significantly different at a 99.9\% confidence level.  For both of these tests, we use injected signals on circular orbits with 100 trial periods from 2-3000 days, 100 values of orbital phase, and with RV amplitudes, $K$, from 1-200\,\ms.  The artificial signals are added to the existing RV data to capture the noise properties of each individual star.  For all stars, we fit and removed any Keplerian signals from confirmed or suspected objects (Table~\ref{tab:fantasyfits}).  The amplitude is increased until the required fraction of signals are deemed detected by the criteria described above.  We test six recovery rates: 99, 90, 70, 50, 30, and 10\%.  This is identical to the approach in our previous work \citep[e.g.][]{periodvalley,witt15,newjupiters} which used the generalised Lomb-Scargle periodogram \citep{zk09} as a detection criterion.

\begin{figure*}
	\includegraphics[width=\columnwidth]{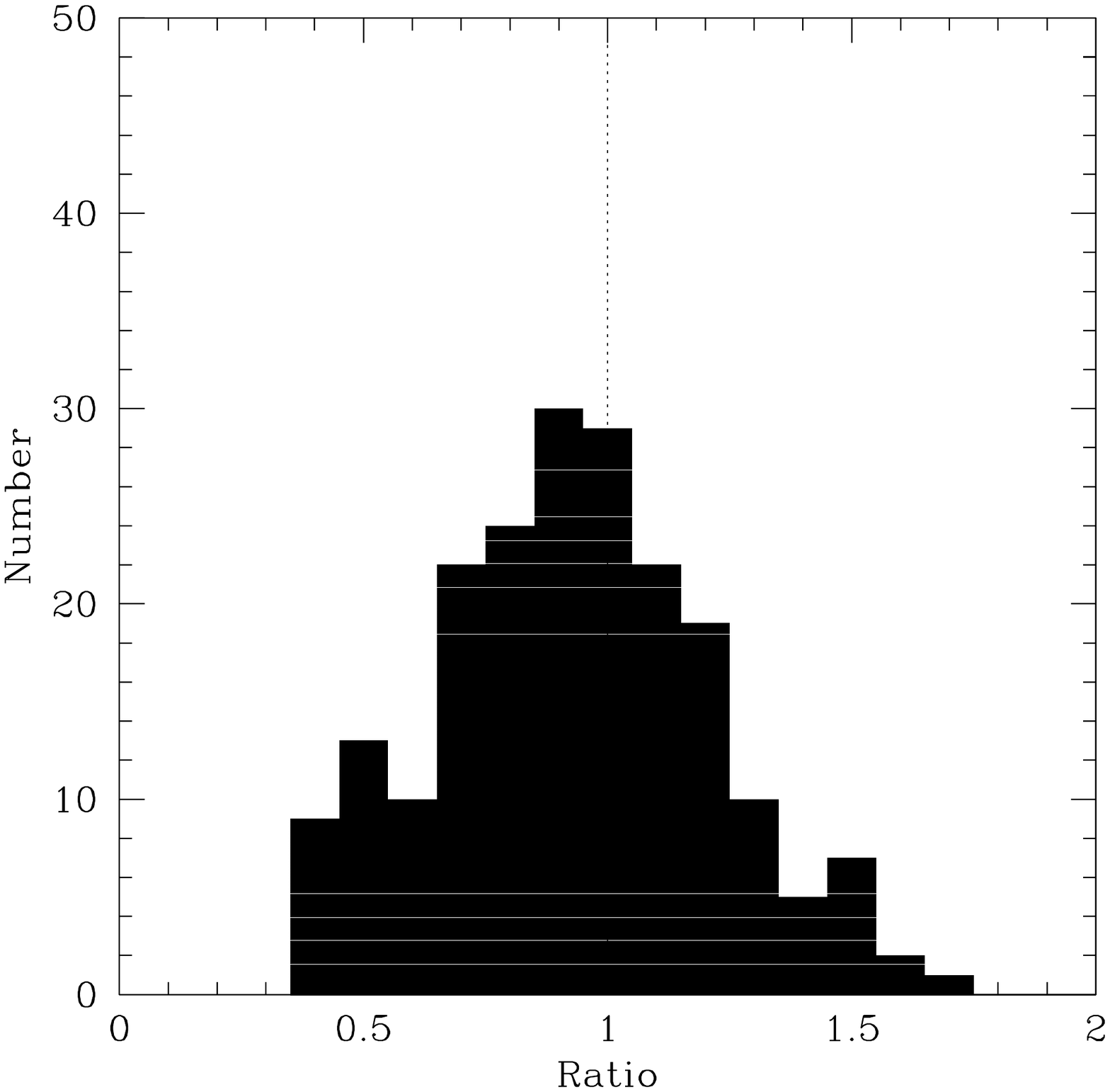}
	\includegraphics[width=\columnwidth]{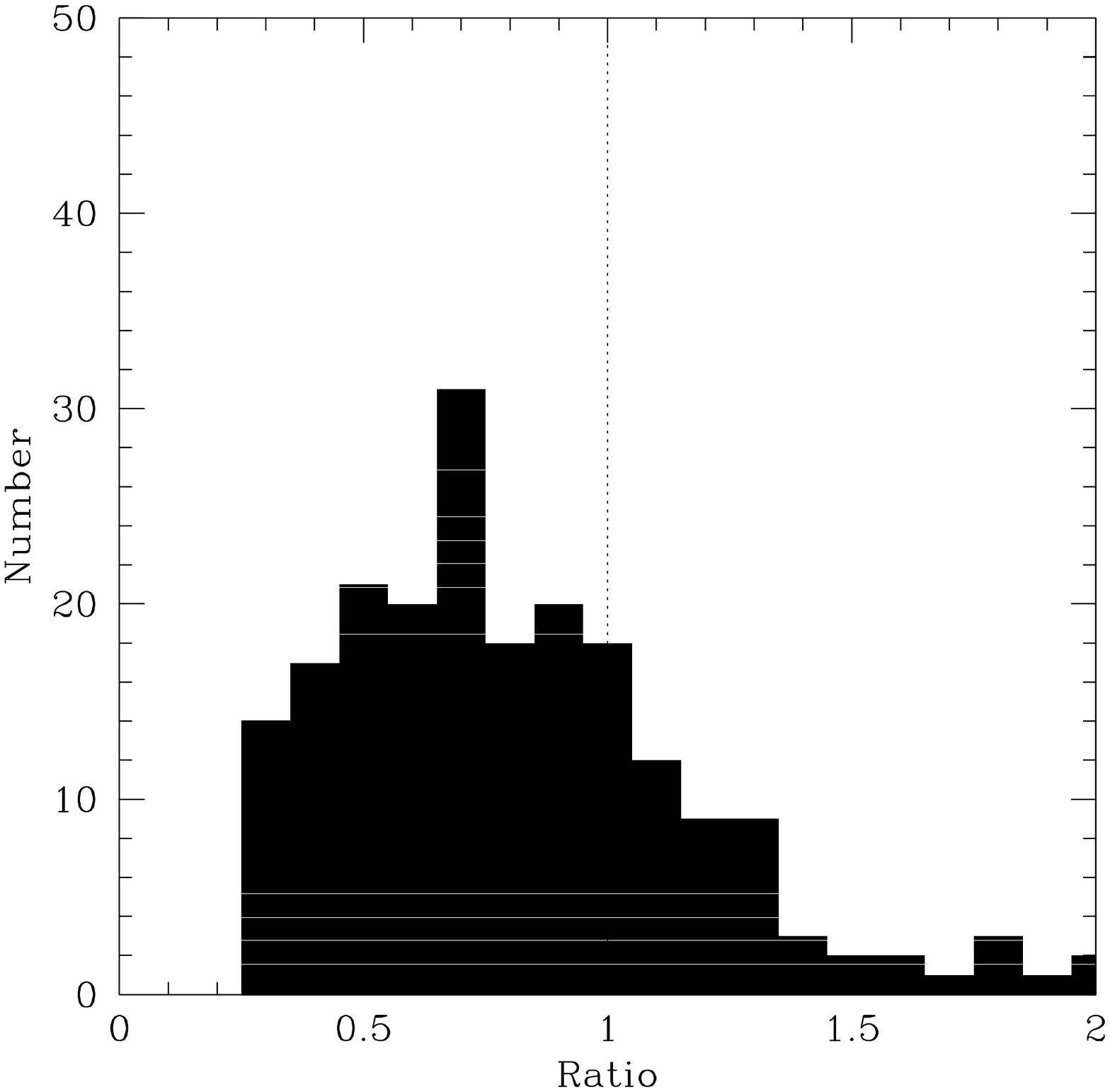}
    \caption{Left: Ratio of mean $K$ detectable from the F-test method versus the periodogram method from our prior work.  Right: Same, but for the RMS method.  The F-test method delivers more consistent results and is adopted for our further analysis. }
    \label{fig:tests}
\end{figure*}

Recognising that these two techniques are quite different from the well-tested periodogram approach of our previous work, we wish to check for any systematic differences between the RMS test and F-test against the ``standard'' detection criterion.  We seek to determine which of these two methods delivers results consistent with the periodogram method.  To do so, we bring both techniques to bear on the Anglo-Australian Planet Search data set that was used in \citet{newjupiters} to assess the occurrence rate of Jupiter analogs.  We use the full RV data set for the 203 stars examined in that work, and apply both the RMS test and F-test to derive detection limits for 100 trial periods between 2-3000 days as described above.  Each trial period produces an RV amplitude that is recovered at the 99\% level.  For each star, we then compute the mean of these 100 RV amplitudes over all periods as the 99\% detection limit $\bar{K}$.  To compare the consistency of the various techniques, we then examine the ratio of $\bar{K}$ as derived from the periodogram test \citep{newjupiters} to the values of $\bar{K}$ obtained for that same RV data set using the RMS and F tests.  Figure~\ref{fig:tests} shows the distribution of those ratios.  As shown in the left panel of Figure~\ref{fig:tests}, the F-test method delivers results that are more consistent, i.e. the distribution is more normal, with a mean ratio of approximately 1.  We therefore adopt the F-test method for all analysis of detection limits in this work.  

\subsection{Occurrence rate of planets around evolved stars}

To determine the underlying occurrence rate of planets around the low-luminosity giants in our sample, we follow the procedure established in our previous work on occurrence rates \citep[e.g.][]{jupiters,etaearth,newjupiters}.  That is, we correct the number of secure detections for the survey incompleteness, to account for planets that may have been missed.  One key difference in this work is that our PPPS sample has some overlap with the EXPRESS survey of \citet{jones11}, and in recent years we have combined efforts to detect planets that our individual data sets could not.  In this section, we consider only those 85 PPPS stars which do not overlap with the EXPRESS targets; the common stars will be considered in a separate analysis (Wolthoff et al., in prep) combined with data from the Lick program \citep{reffert15}.

% this is stolen from Cool Jupiters
Our PPPS-only sample contains observations of 85 stars, from which we have so far confirmed just three  exoplanets; the remaining published discoveries from our data were made with the assistance of data from EXPRESS, and so for the purposes of a uniform sample, we exclude those stars and the planet confirmed in orbit around them.  For each detected planet, we estimate the probability of having detected a planet of that specific period and mass using the results of the injection/recovery simulations described above, summed over the entire sample.  This is accomplished by computing two quantities for each detected planet.  First, for the specific period and mass of the detected planet in question, we calculate the completeness fraction $f_{c}(P,M)$ for the \textit{non-hosts} in the sample:

\begin{equation}
f_{c}(P,M) = \frac{1}{N_{stars}}\sum_{i=1}^{N} f_{R,i}(P,M),
\end{equation}

\noindent where $f_R(P,M)$ is the recovery rate as a function of mass at period $P$, and $N$ is the total number of stars not hosting a planet ($N=82$).  In this way, we account for the detectabilities for each star individually, at each of the 100 trial periods.  The result is the probability that a planet with a given ${P,M}$ would have been detected in the overall sample.  Second, we calculate the recovery rate $f_{R}(P_i,M_i)$ for each detected planet, at the period and mass of that planet.  This represents the probability of having detected that specific planet given the data for that specific star.  These two quantities are then combined in Equation (2) to derive the number of expected detections given the data, and so the number of ``missed'' planets:
%overall occurrence rate of giant planets in a given sample:

\begin{equation}
N_{missed} = \sum_{i=1}^{N_{hosts}} 
\frac{1}{f_{R,i}(P_i,M_i)f_{c}(P_i,M_i)} - N_{hosts}
\end{equation}

\noindent where the symbols have the same meaning as given above.  The occurrence rate of planets in a sample is first estimated as simply the number of detections divided by the total number of stars, using binomial statistics.  The completeness correction in Equation (2) is then used to boost the occurrence rates and their uncertainties by a factor $(N_{missed}+N_{detected})/N_{detected}$ to reflect the imperfect detection efficiency of our observational data.  Applying this procedure to the PPPS-only sample yields a corrected giant planet occurrence rate of 7.8$^{+9.1}_{-3.3}$\% for orbital periods less than about 5 years (i.e. the duration of the PPPS observations).  

% to make the referee happy
We note in passing that before performing the injection-recovery tests, we removed the tentative Keplerian signals presented in Table~\ref{tab:fantasyfits} from the RV data for those 12 stars.  If we do not remove those candidate signals, it is clear that the resulting detection limit will be inflated.  Particularly for the six stellar-mass candidates, the result becomes essentially useless as the scatter of the original data is of order hundreds of \ms.  As per the techniques presented here (Equation 1), those stars then contribute virtually zero detectability information to the sample, and hence the occurrence rates derived from the overall sample will be inflated to reflect the increased number of "missed planets."  The result, in turn, is a higher (but consistent) occurrence rate with larger uncertainties: 9.2$^{+10.8}_{-3.9}$\%.

%\textbf{Jonti: Just a query here - this means 7.8\% of such stars have planets that could be detected by a PPPS type survey, rather than any planets (e.g. Mercury analogues), right? Can we make that more explicit by adding a sentence to the end - something like 'In other words, our sample suggests that $\sim$ 7.8\% of evolved stars are orbited by planets massive enough, and with sufficiently short orbital periods, that a survey with the precision and duration of PPPS would have a reasonable chance of discovering them.}

%-----------------------------------------------------------------------
\section{Summary and Conclusion}

Early estimates of the overall planet occurrence rate for evolved intermediate-mass stars suggested that $\sim$9\% of such stars should host a Jupiter-mass planet \citep{johnson07b}.  The data examined here are not of sufficient quantity or quality to consider the detection of lower-mass planets, and so we restrict our discussion to giant planets ($m$ sin $i \gtsimeq$0.5\,\Mjup) with orbital periods less than 5 years.  Despite this restriction, our result is in broad agreement with that of \citep{johnson07b}, and with the 8.5$\pm$1.3\% giant-planet occurrence rate for such planets orbiting main-sequence stars, as derived by \citet{cumming08}.  

For some time, an observed paucity of giant planets with $a\lesssim$0.5\,au orbiting evolved stars has been a subject of interest \citep[e.g.][]{johnson10, bowler10, weihai}.  The main question has been whether the populations of such close-in planets are different between main-sequence A stars and those ``retired'' A stars as we have examined in the PPPS and other surveys \citep[e.g.][]{villaver09, villaver14, veras16}.  \citet{zhou19} examined the occurrence rates of hot Jupiters orbiting main-sequence AFG stars (spanning the host-star mass range of the PPPS sample: Wittenmyer et al. 2016d), and derived a rate of 0.41$\pm$0.10\%, consistent with the occurrence rate for Solar-type hosts \citep{petigura18, d18}.  Though our PPPS sample is limited, we do achieve relatively high completeness for hot Jupiters ($P<10$ days); with zero detections, the binomial theorem yields an upper limit of 2.7\%, which is in agreement with the result of \citet{zhou19} for main-sequence AFG stars (from which the PPPS population is presumed to have evolved).  

The sample considered here, of the 85 PPPS stars that are not in common with other surveys, contained only three confirmed exoplanet detections.  When the PPPS is considered as a whole, the survey yielded a further 11 planet hosts amongst the the 37 stars in common with the EXPRESS survey.  If we were to include those stars in the analysis described in this work, we would instead derive an overall planet occurrence rate of 31.5$^{+12.2}_{-8.2}$\%, which would be consistent with the \citet{bowler10} result of 26$^{+9}_{-8}$\% resulting from seven detections among 28 subgiant stars.
%It is vaguely distressing that including the Matias stars gives a totally many sigma discrepant occ rate thoough!  
We also note that the next \textit{Gaia} data release, which is expected to include full astrometric orbital solutions, may serve to clarify the nature of the seven large-amplitude signals presented in Table~\ref{tab:fantasyfits}, and may also resolve the mysteries of the sparsely-observed objects in Table~\ref{tab:bigassthings}. 

Taken in concert with other work, our results highlight once again the critical importance of exoplanet surveys with long temporal baselines in driving our understanding of the occurrence of planets moving on long period orbits.  As current and future radial velocity surveys (such as \textsc{Minerva}-Australis) begin to take up the reins from the previous generation (such as the Anglo-Australian Planet Search and the PPPS), and as the astrometric results from \textit{Gaia} become available, we should finally begin to uncover the true diversity of planets moving on longer-period orbits.  Those results will help us to place our own planetary system in context - revealing the presence of Jupiter- and Saturn-analogs, and eventually the abundance of ice giants, like Uranus and Neptune.  By studying evolved stars, and stars both more massive and smaller than our Sun, we will learn the degree to which the Solar system is an unusual product of its environment, or is instead typical of the myriad planetary systems in our galaxy.

% \citet{johnsonoccrate} . 20\%

%    with big enough error bars, everything agrees with everything!!
%but a further five substellar candidates (Table~\ref{tab:fantasyfits}).  If these potential signals are resolved into bona fide planets

% A forthcoming paper (Reffert et al., in prep) will perform a full analysis of the Lick giant-star survey in combination with the EXPRESS survey, totaling XXX stars with considerably larger observational data sets than this work. 

%-----------------------------------------------------------------------
\section*{Acknowledgements}

We acknowledge the traditional owners of the land on which the AAT stands, the Gamilaraay people, and pay our respects to elders past and present.  This material is based upon work supported by the National Science Foundation under Grant No. 1559487 and 1559505.  This research has made use of NASA's Astrophysics Data System (ADS), and the SIMBAD database, operated at CDS, Strasbourg, France.

%%%%%%%%%%%%%%%%%%%%%%%%%%%%%%%%%%%%%%%%%%%%%%%%%%

%%%%%%%%%%%%%%%%%%%% REFERENCES %%%%%%%%%%%%%%%%%%

% The best way to enter references is to use BibTeX:

%\bibliographystyle{mnras}
%\bibliography{example} % if your bibtex file is called example.bib

% Alternatively you could enter them by hand, like this:
% This method is tedious and prone to error if you have lots of references

%%%%%%%%%%%%%%%%%%%%%%%%%%%%%%%%%%%%%%%%%%%%%%%%%%

%%%%%%%%%%%%%%%%% APPENDICES %%%%%%%%%%%%%%%%%%%%%

\appendix

\section{Some extra material}

\begin{table}
%\begin{longtable}
	\centering
	\caption{Complete AAT radial velocity results.  The full version of this table is available online.}
	\label{tab:allvels}
	\begin{tabular}{llrr} 
		\hline
		  Star  & BJD  & RV (\ms)  &  Uncertainty (\ms) \\ 
		\hline
HD100939        &    2454868.10568 &     -97.90 &       2.46 \\ 
HD100939        &    2455969.15781 &      -8.99 &       2.14 \\ 
HD100939        &    2456376.00168 &      12.65 &       2.32 \\ 
HD100939        &    2456399.99654 &      14.99 &       2.24 \\ 
HD100939        &    2456745.08010 &       0.00 &       2.29 \\ 
HD103047        &    2454869.23069 &    -209.69 &       2.24 \\ 
HD103047        &    2455971.08663 &      -9.80 &       1.96 \\ 
HD103047        &    2456059.99580 &      11.14 &       4.41 \\ 
HD103047        &    2456345.08543 &     108.65 &       2.52 \\ 
HD103047        &    2456377.03895 &     115.52 &       2.38 \\ 
		\hline
	\end{tabular}
		\\
%\end{longtable}
\end{table}

% For the love of Odin does anyone know how to make long tables work in MNRAS???
%-----------------------------------------------------------
\begin{table}
	\centering
	\caption{Summary of dispositions for PPPS targets.  Double-lined binary stars (SB2) cannot be used for radial velocity determination, and are reported as having zero observations. }
	\label{tab:dispositions}
	\begin{tabular}{lll} 
		\hline
		  Star  & $N_{obs}$  & Comments  \\ 
		\hline
224910  & 8 &   \\
749  & 0 & SB2   \\
1817 &  14  &   \\
4145 &  9 & Linear trend, $+15.7\pm$0.4 m/s/yr \\
5676 & 6 & Stellar-mass candidate (Table~\ref{tab:bigassthings})  \\
5873 & 0 & SB2 \\
5877 &  0 & SB2  \\
6037 & 14 & Substellar candidate (Table~\ref{tab:fantasyfits}) \\
7931 & 5 &   \\
9218 & 24 &    \\
9925 & 5 &  \\
10731 & 6 &    \\
11343 & 6 & Planet, \citet{jones16}  \\
11653 & 3 & Stellar-mass candidate (Table~\ref{tab:bigassthings})  \\
12974 & 3 &  \\
13471 & 7 &   \\
13652 & 8 & Substellar candidate (Table~\ref{tab:fantasyfits}) \\
14805 & 6 &   \\
14791 & 4 & Stellar-mass candidate (Table~\ref{tab:bigassthings})  \\
15414 & 5 &   \\
19810 & 3 &   \\
20035 & 0 & SB2  \\
20924 & 13 &   \\
24316 & 7 &   \\
25069 & 15 &  \\
28901 & 15 &   \\
29399 & 22 & Strong activity cycle, \citet{witt17a}  \\
31860 & 0 & SB2  \\
34851 & 9 & Binary, \citet{ppps3} \\
33844 & 20 & Planets, \citet{ppps4} \\
37763 & 20 & Stellar-mass candidate (Table~\ref{tab:fantasyfits})  \\
39281 & 13 &  \\
40409 & 27 & Linear trend, $-23.0\pm$0.2 m/s/yr  \\
43429 & 20 & Stellar-mass candidate (Table~\ref{tab:fantasyfits})  \\
46122 & 0 & SB2 \\
46262 & 16 &  \\
47141 & 14 &  \\
47205 & 27 & Planet, \citet{47205}  \\
51268 & 16 & Stellar-mass candidate (Table~\ref{tab:bigassthings})  \\
58540 & 0 & SB2   \\
59663 & 11 &   \\
67644 & 12 &    \\
72467 & 12 &    \\
76321 & 0 & SB2   \\
76437 & 15 &  Linear trend, $+6.5\pm$0.3 m/s/yr  \\
76920 & 17 & Planet, \citet{76920}  \\
80275 & 8 & Linear trend, $+14.6\pm$0.4 m/s/yr  \\
81410 & 0 & SB2    \\
84070 & 7 & Stellar-mass candidate (Table~\ref{tab:bigassthings})  \\
85128 & 8 &   \\
85035 & 24 &   \\
86359 & 6 &  \\
87089 & 9 & Quadratic trend  \\
86950 & 20 & Planet, \citet{witt17a}  \\
HIP50638 & 9 &   \\
94386 & 14 & Binary, \citet{ppps3}  \\
95900 & 10 &  \\
98516 & 16 &   \\
98579 & 0 & SB2  \\
100939 & 5 & Planet, \citet{jones19}  \\
103047 & 5 &  \\
104358 & 12 & Binary, \citet{ppps3}  \\
104704 & 6 & Linear trend, $+9.2\pm$0.4 m/s/yr   \\
104819 & 3 & Stellar-mass candidate (Table~\ref{tab:bigassthings})  \\
105096 & 9 & Quadratic trend  \\
105811 & 9 & Binary, \citep{bluhm16}  \\
106314 & 11 &  \\
108991 & 16 &   \\
109866 & 9 &  \\
110238 & 5 & Stellar-mass candidate (Table~\ref{tab:bigassthings})  \\
114899 & 11 & Substellar candidate (Table~\ref{tab:fantasyfits})  \\
115066 & 15 & Stellar-mass candidate (Table~\ref{tab:fantasyfits})  \\
115202 & 20 &  \\
117434 & 3 &  \\
121056 & 19 & Planets, \citet{121056}  \\
121156 & 10 & Stellar-mass candidate (Table~\ref{tab:fantasyfits})  \\
121930 & 10 &   \\
124087 & 7 & Stellar-mass candidate (Table~\ref{tab:bigassthings})  \\
125774 & 7 &  \\
126105 & 15 & Substellar candidate (Table~\ref{tab:fantasyfits})  \\
130048 & 13 &   \\
131182 & 5 &   \\
132396 & 16 &   \\
133166 & 0 & SB2  \\
133670 & 16 &   \\
134443 & 7 &   \\
134692 & 7 &   \\
135760 & 13 & Planet, \citet{jones16}  \\
135872 & 3 & Linear trend, $+40.5\pm$1.4 m/s/yr  \\
136295 & 15 & Planet, \citet{jones19}  \\
137115 & 3 &   \\
137164 & 0 & SB2 \\
136135 & 5 & Quadratic trend  \\
138061 & 4 &  \\
138716 & 18 &   \\
138973 & 4 & Quadratic trend  \\
142132 & 6 & Stellar-mass candidate (Table~\ref{tab:fantasyfits})  \\
142384 & 0 & SB2  \\
143561 & 5 &   \\
144073 & 7 &   \\
145428 & 9 & Stellar-mass candidate, \citet{luhn19} and Table~\ref{tab:fantasyfits} \\
148760 & 13 & Quadratic trend  \\
153438 & 0 & SB2  \\
154250 & 7 &   \\
155233 & 21 & Planet, \citet{ppps3}  \\
154556 & 12 &   \\
159743 & 10 & Substellar candidate (Table~\ref{tab:fantasyfits})  \\
162030 & 20 &   \\
166309 & 5 & Stellar-mass candidate (Table~\ref{tab:bigassthings})  \\
166476 & 4 & Linear trend, $+15.1\pm$0.5 m/s/yr  \\
170707 & 4 & Planet, \citet{jones19}  \\
170286 & 8 &   \\
173902 & 16 & Quadratic trend   \\
176002 & 10 &   \\
175304 & 4 &    \\
177897 & 7 &  \\
176794 & 0 & SB2  \\
181342 & 5 & Planet, \citet{jones16}  \\
181809 & 7 & Stellar-mass candidate (Table~\ref{tab:bigassthings})  \\
188981 & 16 & Binary, \citet{ppps3}  \\
191067 & 6 &   \\
196676 & 6 & Quadratic trend  \\
199809 & 4 & Linear trend, $+23.0\pm$0.4 m/s/yr  \\
200073 & 14 & Linear trend, $+74.1\pm$0.3 m/s/yr  \\
201931 & 11 &   \\
204073 & 11 &  \\
204057 & 3 & Stellar-mass candidate (Table~\ref{tab:bigassthings})  \\
204203 & 0 & SB2   \\
205577 & 8 & Substellar candidate (Table~\ref{tab:fantasyfits})  \\
205972 & 7 &   \\
208431 & 6 &  \\
208791 & 4 &    \\
208897 & 3 &  \\
214573 & 12 &   \\
216640 & 21 &   \\
218266 & 6 &   \\
219553 & 9 & Planet, \citet{jones19}  \\
222076 & 11 & Planet, \citet{witt17a}  \\
222768 & 3 & Stellar-mass candidate (Table~\ref{tab:bigassthings})  \\
223301 & 5 &   \\
223860 & 4 &   \\
		\hline
	\end{tabular}
		\\
\end{table}

%%%%%%%%%%%%%%%%%%%%%%%%%%%%%%%%%%%%%%%%%%%%%%%%%%

% Don't change these lines
\bsp	% typesetting comment
\label{lastpage}
\end{document}